\documentclass{PoS}

\usepackage{amsmath}
\usepackage{cite}
\usepackage{verbatim}

\title{Can a light Higgs impostor hide in composite gauge models?}

\ShortTitle{Can a light Higgs impostor hide in composite gauge models?}

\author{Zolt\'{a}n Fodor\\
        University of Wuppertal\\
        E-mail: \email{fodor@bodri.elte.hu}}

\author{Kieran Holland\\
        University of the Pacific\\
		Albert Einstein Center for Fundamental Physics, Institute for
        Theoretical Physics, Bern University\\
        E-mail: \email{kholland@pacific.edu}}

\author{Julius Kuti\\
        University of California, San Diego\\
        E-mail: \email{jkuti@ucsd.edu}}

\author{D\'{a}niel N\'{o}gr\'{a}di\\
        E\"{o}tv\"{o}s University\\
        E-mail: \email{nogradi@bodri.elte.hu}}

\author{\speaker{Chik Him Wong}\\
        University of California, San Diego\\
        E-mail: \email{rickywong@physics.ucsd.edu}}

\abstract{
	The frequently discussed strongly interacting gauge theory with a fermion flavor doublet in the two-index symmetric (sextet) representation of the SU(3) color gauge group is investigated\cite{hong2004,dietrich2005,plb2012}. In previous studies\cite{plb2012} the chiral condensate and the mass spectrum were shown to be consistent with chiral symmetry breaking ($\chi$SB) at vanishing fermion mass. The recently reported $\beta$-function \cite{degrand2012} is not inconsistent with this observation, suggesting that the model is very close to the conformal window and a light ``Higgs impostor'' could emerge as a composite state. In this work we describe the methodology and preliminary results of studying the emergence of the light composite scalar with $0^{++}$ quantum numbers.
}

\FullConference{31st International Symposium on Lattice Field Theory - LATTICE 2013\\
		July 29 - August 3, 2013\\
		Mainz, Germany}

\begin{document}

\section{Introduction}

The newly discovered Higgs-like particle with decay modes close to that of the Standard Model brings new focus to the search for theoretical frameworks. An example is the light composite scalar as a viable interpretation of the discovery. Nearly conformal gauge theories serve as the theoretical laboratories for credible realization of such a scenario\cite{hong2004,dietrich2005,plb2012,yamawaki1986,bardeen1986,holdem1988,goldberger2008,appelquist2010,grinstein2011,antipin2012,hashimoto2011,matsuzaki2012}. In this work we investigate the minimal realization with a fermion flavor doublet in the two-index symmetric (sextet) representation of the SU(3) color gauge group (reviewed in \cite{nogradi2012a,kuti2013}) close to the conformal window. The smallness of the $\beta$-function \cite{degrand2012} implies either the existence of a conformal fixed point or a slowly walking scenario\cite{kogut2010,kogut2011}. Consistency with $\chi$SB at vanishing fermion mass, reported in \cite{plb2012,holland2013}, would require the sextet model to remain just below the conformal window with a very small but non-vanishing $\beta$-function. As suggested by \cite{hong2004,dietrich2005,plb2012}, this model with the simplest composite Higgs mechanism leaves open the possibility of a light scalar state, possibly as the dilaton state of broken scale invariance. With or without the dilaton interpretation, such a state could serve as a Higgs impostor. In this work we report preliminary results in an attempt to address these important problems.

\section{Methodology}
The quantum numbers of the $f_0$ meson match those of the $0^{++}$ state. Close to the conformal window, the $f_0$ meson is not expected to be similar to the counterpart in QCD. If it turns out to be light, it could replace the role of the elementary Higgs particle.  The two types of $0^{++}$ operators, the fermionic one and the gluonic one ($0^{++}$ glueball), are expected to mix in the ground state. Such mixing is ignored in this work and will be discussed in future reports. Only fermionic operators are discussed here.\\
\ \\
Flavor-singlet fermionic correlators have fermion-line connected and fermion-line disconnected contributions from fermion loop diagrams. The latter one is often known as the disconnected diagram and denoted by $D(t)$ at time separation $t$. The connected diagram is the same as that of the non-singlet correlator and is denoted as $C_{\rm non-singlet}(t)$. 
The $f_0$ correlator $C_{\rm singlet}(t)$ is defined as $C_{\rm singlet}(t) \equiv C_{\rm non-singlet}(t) + D(t)$. The positive definite nature of the transfer matrix guarantees the spectral decomposition of $C_{\rm singlet}(t)$ in terms of the energy levels $m_i^{0^{++}}$ with the parity partners $m_j^{0^{-+}}$ for staggered fermions. On a lattice with temporal extent $T$, 
\begin{align}\label{decompose}
	C_{\rm singlet}(t) &= \sum_i b_i \cosh(m_i^{0^{++}}(T/2-t)) + (-)^t \sum_j b'_j \cosh(m_j^{0^{-+}}(T/2-t)) + {\rm vev}^2\\\notag
	&\approx b_0 \cosh(m_0^{0^{++}} (T/2-t)) + (-)^{t} b'_0 \cosh(m_0^{0^{-+}}(T/2-t)) + {\rm vev}^2
\end{align}
at large $t$, where $m_0^{0^{++}}$ and $m_0^{0^{-+}}$ correspond to $m_{f_0}$ and $m_{\eta_{\rm SC}}$ respectively . To evaluate the disconnected diagram, one needs to calculate quark propagators that start and end on the same spacetime site. These propagators are stochastically estimated to avoid costly $O(V)$ inversions. Both connected and disconnected diagrams are estimated by these propagators in the staggered formalism.
We introduce $Z_2$ noise sources on the lattice as follows. Each source is defined on individual time-slice $t_0$ and color $a$. The ones defined on even spatial sites are denoted by $\eta_{[E]}^a(t_0)$ and the ones defined on odd sites are denoted by $\eta_{[O]}^a(t_0)$. It can be viewed as a ``dilution'' scheme which is fully diluted in time and color and even/odd diluted in space\cite{foley2005}. The corresponding quark propagators with destination site $(\vec{x}_1,t_1)$ are defined as $\varphi_{[E,a,t_0,U]}(\vec{x}_1,t_1)$ and $\varphi_{[O,a,t_0,U]}(\vec{x}_1,t_1)$ respectively. The diagrams are then computed from the relations
\begin{align}
	& C_{\rm non-singlet}(t) =(-)^{t+1+\vec{x}_1 \cdot \vec{n}} \cdot \\\notag
	&\langle {\rm Tr} (\varphi_{[E,a,t_0,U]}(\vec{x}_1,t_0+t) \varphi^\dagger_{[E,a,t_0,U]}(\vec{x}_1, t_0+t) - \varphi_{[O,a,t_0,U]}(\vec{x}_1, t_0+t) \varphi^\dagger_{[O,a,t_0,U]}(\vec{x}_1,t_0+t))\rangle_{U,t_0,\eta}, \\\notag
	&D(t) =\frac{N_f}{4} m^2 \cdot \\\notag
	& \langle {\rm Tr} (\varphi_{[E,a,t_0+t,U]}(\vec{x}_1,t_0+t) \varphi^\dagger_{[E,a,t_0+t,U]}(\vec{x}_1, t_0+t) + \varphi_{[O,a,t_0+t,U]}(\vec{x}_1,t_0+t) \varphi^\dagger_{[O,a,t_0+t,U]}(\vec{x}_1,t_0+t)) \cdot \\\notag
	\ & {\rm Tr} (\varphi_{[E,a,t_0,U]}(\vec{x}_2,t_0) \varphi^\dagger_{[E,a,t_0,U]}(\vec{x}_2,t_0) + \varphi_{[O,a,t_0,U]}(\vec{x}_2,t_0) \varphi^\dagger_{[O,a,t_0,U]}(\vec{x}_2, t_0)) \rangle_{U,t_0,\eta},
\end{align}
in which $N_f=2$, $U$ designates the gauge links, $\vec{n}=(1,1,1)$, and the identity ${\rm Tr} M^{-1}=m{\rm Tr} (M^\dagger M)^{-1}$ is used, in which $m$ is the fermion mass and $M$ is the Dirac matrix in the staggered formalism. In our calculation different noise sources are used for each choice of index E/O, $a$, $t_0$ and each gauge configuration.
\begin{figure}
	\begin{center}
	\scalebox{0.195}{\includegraphics{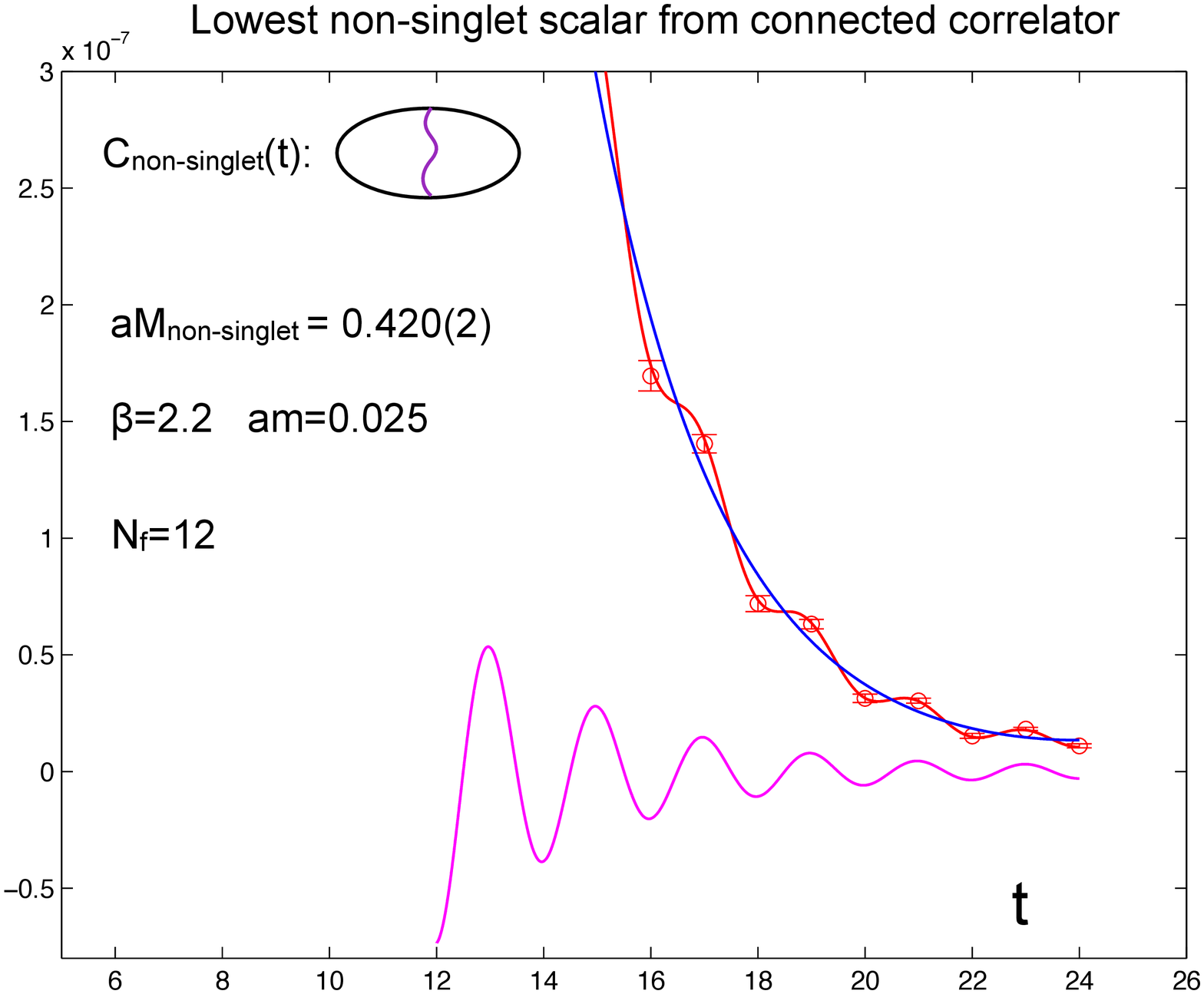}}
	\scalebox{0.185}{\includegraphics{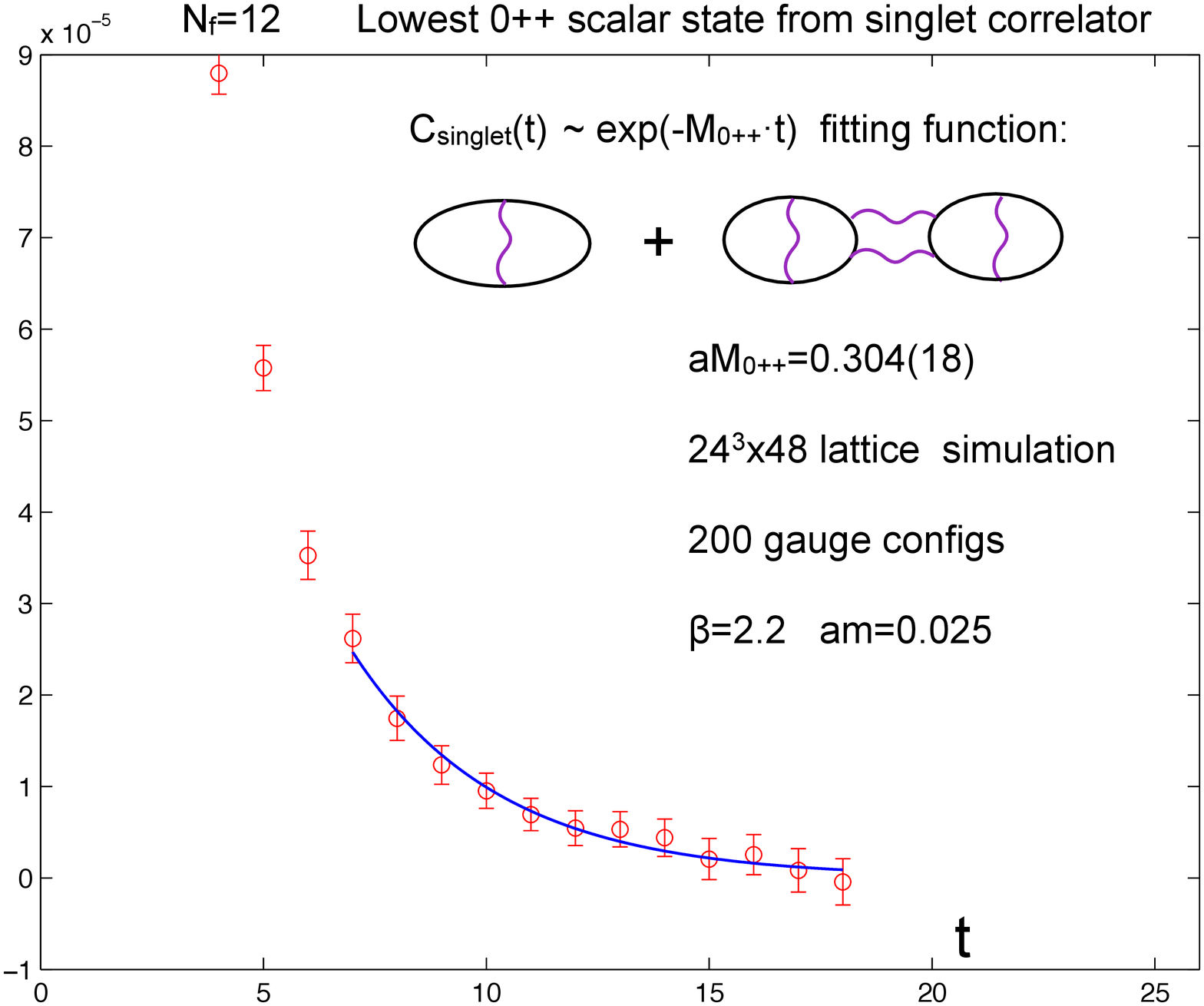}}
	\scalebox{0.29}{\includegraphics{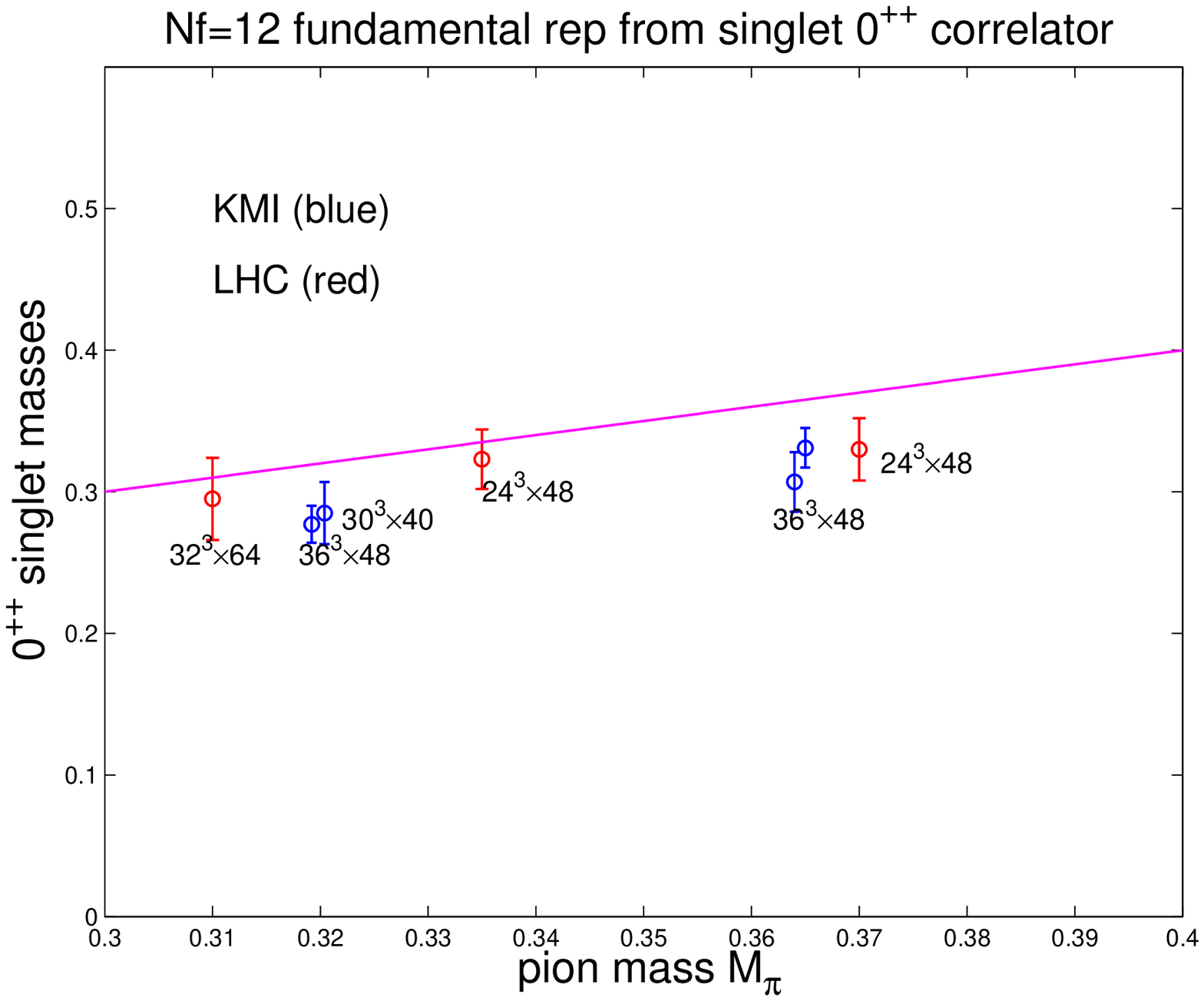}}
	\caption{Test on $N_f=12$ fundamental SU(3) model. The tree-level Symanzik-improved gauge action with staggered fermions\cite{aoki2006} is used and the HMC algorithm is employed. Here we choose $\beta=2.20$ and the fermion mass $a m=0.025$. We use $200$ gauge configurations separated by $20$ MD time units. (Left) A typical non-singlet correlator (red points) has a non-oscillating contribution from $a_0$ (fitted as blue curve) and an oscillating contribution from the parity partner $\pi_{\rm sc}$ (fitted as magenta curve). The red curve is the combination of the two fits. (Center) A typical singlet correlator has a non-oscillating contribution from $f_0$ (fitted as blue curve) and the oscillating contribution $\eta_{\rm SC}$ is not detectable within statistical fluctuation. (Right) Our results agree with results of latKMI collaboration\cite{kmi2013} within comparable accuracy\cite{kuti2013}. The magenta curve indicates the $m_{\pi}$ value for comparison.}\label{kmi} 
	\end{center}
\end{figure}
Fig. \ref{kmi} shows a test of this method performed in the $N_f=12$ fundamental SU(3) model, which is either conformal or nearly conformal. Our results are consistent with results the latKMI group obtained using similar techniques within comparable accuracies\cite{kmi2013}. The $f_0$ mass is found to be light in the $N_f=12$ model, providing first evidence that a light $0^{++}$ state can emerge in strongly interacting gauge models.  

\section{Spectroscopy Analysis}
\begin{figure}
	\begin{center}
		\scalebox{0.25}{\includegraphics{sample/a04.eps}}
		\scalebox{0.25}{\includegraphics{sample/f0-a03.eps}}
		\scalebox{0.5}{\includegraphics{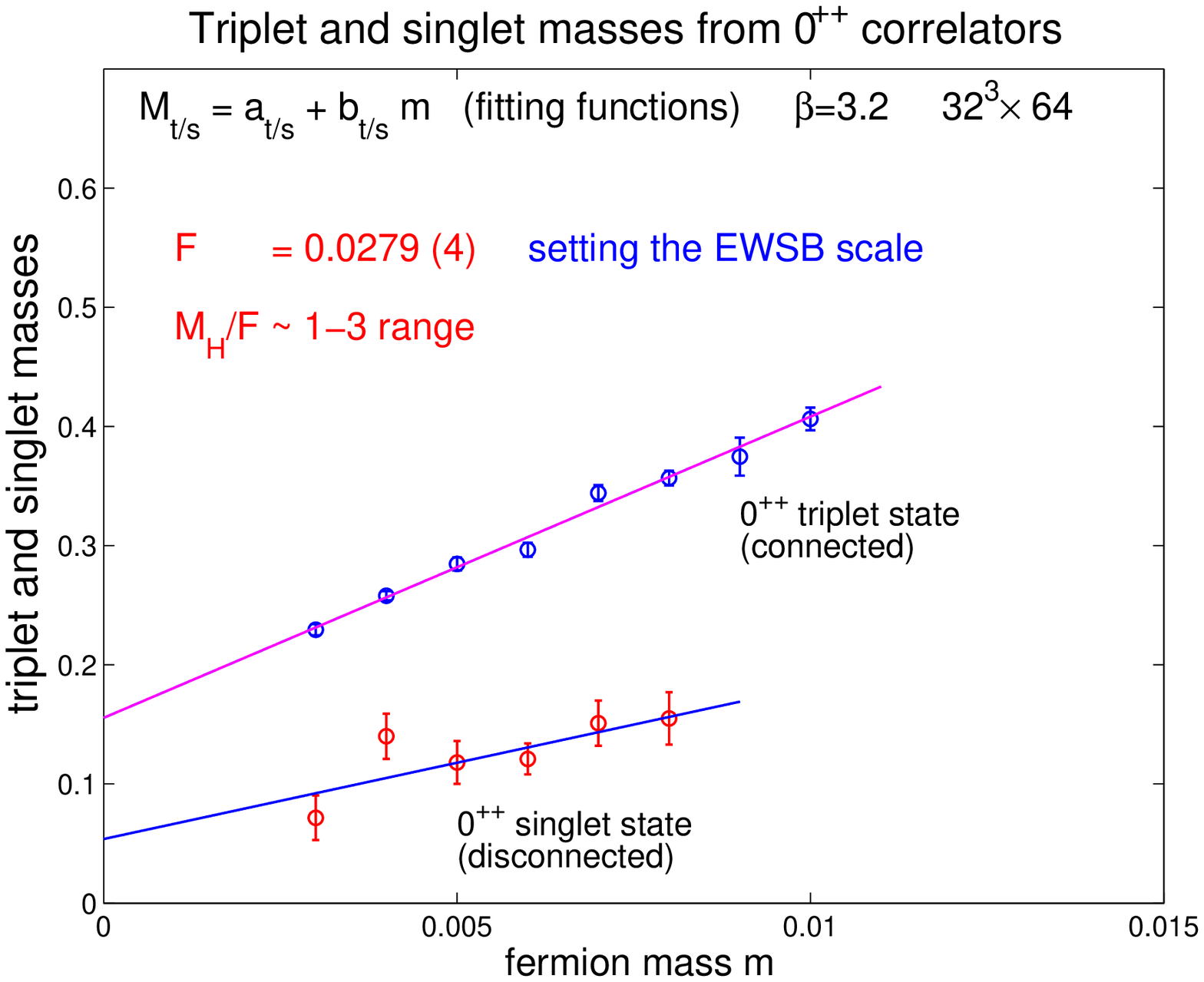}}
		\caption{Preliminary results on $32^3 \times 64$ lattices at $\beta=3.20$. (Top left) A typical non-singlet correlator can be fitted well with non-oscillating $a_0$ contribution and oscillating $\pi_{\rm SC}$ contribution. (Top right) A typical $\tilde{D}(t)$ can be fitted well with a single $m_D$. This can be identified as $m_{f_0}$ as explained in text. No oscillating contribution is detectable within errors. (Bottom) Preliminary $f_0$ masses at different fermion masses. Higher statistics and more comprehensive analysis are required for a more robust extrapolation to the chiral limit.}\label{analysis}
	\end{center}
\end{figure}

Our simulations use the tree-level Symanzik-improved gauge action with two-flavor staggered fermions in the sextet representation of the SU(3) gauge group. The RHMC algorithm is employed. For molecular dynamics time evolution we applied multiple time scales and the Omelyan integrator. Fig. \ref{analysis} shows the preliminary results at $\beta\equiv6/g^2=3.20$.  There are $135$ gauge configurations separated by $20$ MD time units. Autocorrelations are monitored by the time histories of effective masses and correlators. 
The top left plot shows a typical non-singlet correlator, which can be fitted well by the ansatz:\\
\begin{equation}
	C_{\rm non-singlet}(t) = c_0 (\cosh(m_{a_0}(T/2-t)) + (-)^{t} c_1 \cosh(m_{\pi_{\rm{SC}}}(T/2-t)))
\end{equation}
as expected. Here $m_{a_0}$ is the $a_0$ mass, $m_{\pi_{\rm SC}}$ is the mass of the parity partner. Since the ${\rm vev}^2$ of $D(t)$ is irrelevant, we analyze the subtracted disconnected diagram $\tilde{D}(t)\equiv D(t)-D(T/2)$(top right plot). Empirically no oscillation is detected within error. It can be fitted well with the ansatz:\\
\begin{equation}\label{fit}
	\tilde{D}(t) = c_0(\cosh(m_D(T/2-t))-1),
\end{equation}
with $m_D < m_{a_0}$ as a general empirical observation. 
Comparing Eq. \ref{fit} and \ref{decompose} in the large $t$ limit, the identification $m_{f_0} \approx m_{D}$ can be made empirically, as long as $m_D < m_{a_0}$. In other words, $m_D$ from fitting $\tilde{D}(t)$ alone provides the estimate of the lowest state of $0^{++}$ that couples to the fermionic operator. This state can be identified as $m_{f_0}$ instead of other excited levels, since it is reasonable to assume that the ground state of $0^{++}$ overlaps with the fermionic operator. A more comprehensive $0^{++}$ spectroscopy including the mixing with gluonic operators requires careful variational analysis and is beyond the scope of this work.\\

We define a three-point effective mass $m_{\rm eff}$ as follows:
\begin{align}
			&\frac{\tilde{D}(t)+2\tilde{D}(t+1)+\tilde{D}(t+2)}{\tilde{D}(t-1)+2\tilde{D}(t)+\tilde{D}(t+1)}\\\notag
			\equiv&  \frac{\cosh(m_{\rm{eff}} (T/2-t)) + 2\cosh(m_{\rm{eff}}(T/2-(t+1)))  + \cosh(m_{\rm{eff}} (T/2-(t+2)))-4}{\cosh(m_{\rm{eff}} (T/2-(t-1)))+2\cosh(m_{\rm{eff}} (T/2-t)) +\cosh(m_{\rm{eff}} (T/2-(t+1)))-4}
\end{align}
and fit it as a constant over a selected $t$-range. Thermalization is monitored by ensuring stability of the fitted masses along the trajectory. The autocorrelation among gauge configurations is reduced by measuring on well-separated configurations. A principle component analysis is performed by discarding very small eigenvalues or eigenvalues with too large relative errors in covariance matrices. The bottom plot of Fig. \ref{analysis} shows the preliminary $f_0$ masses at different fermion masses, in comparison with the $a_0$ masses. The analysis uses configurations separated by $20$ MD time units. Although much higher statistics is required for more robust extrapolation to the chiral limit, it is clear that $m_{f_0}$ is much lower than $m_{a_0}$ and, in the chiral limit, $m_{0^{++}} \sim (1-3)F$,  translating into a range of $250$ to $750$ GeV.  According to \cite{foadi2012}, this mass range for $f_0$ is sufficiently low to be further downshifted by the top quark loop self energy to make it compatible with the experimentally-observed Higgs state.\\
\begin{figure}
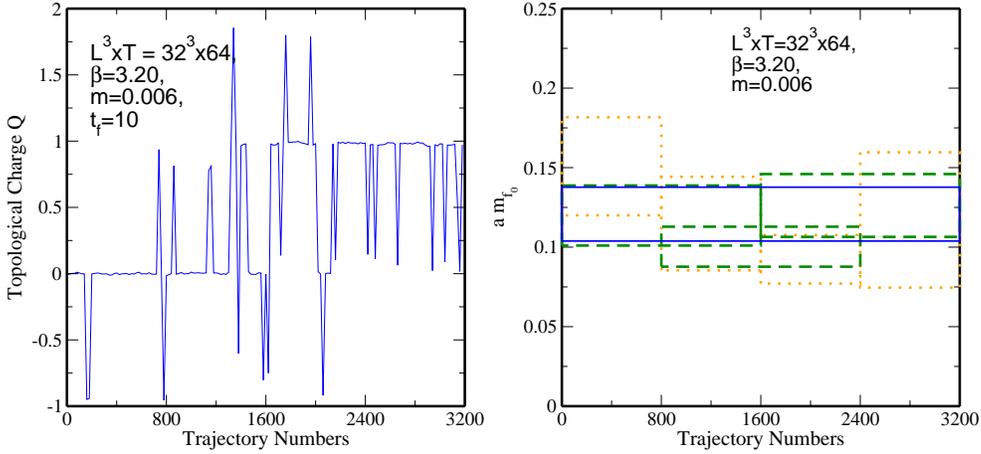

	\begin{center}
	\scalebox{0.35}{\includegraphics{m0p006/wt10p0data/top.eps}}
	\scalebox{0.35}{\includegraphics{top/plot.eps}}
	\caption{Investigating any topological charge $Q$ dependence of fitted $m_{f_0}$ for different segments along the RHMC trajectory. (Left) Topological charge history along the trajectory. $Q$ is measured at gradient flow step $t_f=10$ at $c=0.3$ \cite{nogradi2012}. (Right) Fitted values of $m_{f_0}$ using gauge configurations indicated by the horizontal extents of the boxes. The configurations are separated by $20$ MD time units. Boxes of the same color contain the same number of configurations. It is observed that although $Q$ changes slowly along the trajectory, the fitted $m_{f_0}$ remains statistically the same. This may indicate that the dependence is insignificant. However a reliable conclusion requires a more systematic study with higher statistics. Possible effects from thermalization and autocorrelation should also be taken into account.}\label{top}
\end{center}
\end{figure}

Although the fitting strategies described above are reasonable and serve well the purpose of the preliminary study, there are two main issues to be addressed for reliable results. It is well-known that along the Markov chain, the tunneling of topology is slow. The sensitivity of the fitted mass values of $f_0$ to the slowly changing topology is still not clear. According to our preliminary data shown in Fig. \ref{top} , such effects seem to be insignificant. However, when higher statistics are available, this needs to be investigated more carefully. The existence of a light $f_0$ state makes the extrapolation to vanishing fermion mass difficult. The low-lying $f_0$ state will interact with the pion and hence requires the modification of $\chi$-PT. 

\section{Conclusion and Outlooks}
The $0^{++}$ state in the sextet representation of the SU(3) model with $N_f=2$ fermions is studied. Stochastic estimation is employed and only the fermionic operator is used. A light $f_0$ mass is obtained as a preliminary result, providing the first necessary step for the realization of the composite Higgs mechanism with a light Higgs particle. While the preliminary results are encouraging, there are a few important issues to be addressed in the analysis. In addition to topological effects and modified $\chi$-PT, the mixing between fermionic and gluonic operators is also ignored in this work. If the lowest $0^{++}$ state overlapping with the gluonic operator is light near the conformal window, such mixing in the ground state is expected to be significant. This will be investigated by a detailed variational analysis. Also, a careful analysis of finite volume and cutoff effects is required to extrapolate to infinite volume and continuum limits. All these studies are ongoing and will be reported in future publications.

\section{Acknowledgements}
We acknowledge support by the DOE under grant DE-SC0009919, by the NSF under grants 0704171 and 0970137, by the EU Framework Programme 7 grant (FP7/2007-2013)/ERC No 208740, by OTKA under the grant OTKA-NF-104034, and by the Deutsche Forschungsgemeinschaft grant SFB-TR 55. Computational resources were provided by USQCD at Fermilab and JLab, by the NSF XSEDE program, and by the University of Wuppertal. CUDA port of the simulation code is provided by the authors of \cite{gyozo2007}. KH wishes to thank the Institute for Theoretical Physics and the Albert Einstein Center for Fundamental Physics at Bern University for their support. KH and JK wish to thank the Galileo Galilei Institute for Theoretical Physics and INFN for their hospitality and support at the workshop ``New Frontiers in Lattice Gauge Theories''.


\begin{thebibliography}{99}

	\bibitem{hong2004} D.~K.~Hong, S.~D.~H.~Hsu and F.~Sannino, Phys.\ Lett.\ B {\bf 597}, 89 (2004).
	\bibitem{dietrich2005} D.~D.~Dietrich, F.~Sannino and K.~Tuominen, Phys.\ Rev.\ D {\bf 72}, 055001 (2005).
	\bibitem{plb2012} Z. Fodor,K. Holland, J. Kuti, D. Nogradi, C. Schroeder and C. H. Wong, Phys Lett. B {\bf 718},657 (2012).
	\bibitem{degrand2012} T. DeGrand, Y. Shamir and B. Svetitsky, Phys. Rev. D {\bf 87}, 074507 (2013).
	\bibitem{yamawaki1986}K. Yamawaki, M. Bando, K. -i. Matumoto, Phys. Rev. Lett. {\bf 56}, 1335 (1986), 
	\bibitem{bardeen1986} W. A. Bardeen, C. N. Leung and S. T. Love, Phys. Rev. Lett. {\bf 56}, 1230 (1986).
	\bibitem{holdem1988} B. Holdom and J. Terning, Phys. Lett. B {\bf 187}, 357 (1987); Phys. Lett. B {\bf 200}, 338 (1988).
	\bibitem{goldberger2008}W. D. Goldberger, B. Grinstein and W. Skiba, Phys. Rev. Lett.{\bf 100}, 111802 (2008).
	\bibitem{appelquist2010} T. Appelquist and Y. Bai, Phys. Rev. D {\bf 82}, 071701 (2010).
	\bibitem{grinstein2011} B. Grinstein and P. Uttayarat, JHEP {\bf 1107}, 038 (2011).
	\bibitem{antipin2012} O. Antipin, M. Mojaza and F. Sannino, Phys. Lett. B {\bf 712}, 119 (2012).
	\bibitem{hashimoto2011} M. Hashimoto and K. Yamawaki, Phys. Rev. D {\bf 83}, 015008 (2011).
	\bibitem{matsuzaki2012} S. Matsuzaki and K. Yamawaki, Phys. Rev. D {\bf 86},035025 (2012).
	\bibitem{nogradi2012a} D. Nogradi, PoS LATTICE2011 010 (2011).
	\bibitem{kuti2013} J. Kuti, PoS LATTICE2013 004 (2013).
	\bibitem{kogut2010} J.B. Kogut and D. K. Sinclair, Phys. Rev. {\bf D81},114507(2010).
	\bibitem{kogut2011} J.B. Kogut and D. K. Sinclair, Phys. Rev. D {\bf 84},074504(2011).
	\bibitem{holland2013} Z. Fodor,K. Holland, J. Kuti, D. Nogradi, C. Schroeder and C. H. Wong, PoS LATTICE2013, 089(2013).
	\bibitem{foley2005} J. Foley, K. Juge, A. Cais, M. Peardon, S. Ryan and J. Skullerud, Comput. Phys. Commun. {\bf 172}, 145-162 (2005).
	\bibitem{aoki2006} Y. Aoki, Z. Fodor, S.D. Katz and K.K. Szabo, JHEP01, 089 (2006)
	\bibitem{kmi2013} Y. Aoki,T. Aoyama, M. Kurachi, T. Maskawa, K. Nagai, H. Ohki, E. Rinaldi, A. Shibata, K. Yamawaki and T. Yamazaki, Phys. Rev. Lett. {\bf 111} 162001(2013).
	\bibitem{foadi2012} R. Foadi, M. Frandsen, F. Sannino, Phys. Rev. D {\bf 87} 095001 (2013)
	\bibitem{nogradi2012} Z. Fodor,K. Holland, J. Kuti, D. Nogradi, C. Schroeder and C. H. Wong, PoS LATTICE2012,050 (2012).
	\bibitem{gyozo2007} G. Egri, Z. Fodor, C. Hoelbling, S. Katz, D. Nogradi and K. Szabo, Comput. Phys. Commun. {\bf 177} 631-639 (2007).
\end{thebibliography}
\end{document}